\documentclass[aps,pre,preprint,showpacs,amsmath,preprintnumbers]{revtex4}
\usepackage{epsfig}
\newcommand{\Op}[1]{{\boldsymbol{\mathrm{\hat{#1}}}}}
\begin{document}

\title{On the temperature dependence of the interaction-induced entanglement.}
\author{Michael Khasin and Ronnie Kosloff }
\affiliation{Fritz Haber Research Center for Molecular Dynamics, Hebrew University of Jerusalem, Jerusalem 91904,
Israel}
\date{\today }

\begin{abstract}
Both direct and indirect weak nonresonant interactions are shown to produce entanglement 
between two initially disentangled systems prepared as a tensor product 
of thermal states, provided the initial temperature is sufficiently low. 
Entanglement is determined by the Peres-Horodeckii criterion, which establishes that a composite state is entangled
if its partial transpose is not positive. 
If the initial temperature of the thermal states is higher than an upper 
critical value $T_{uc}$ the minimal eigenvalue of the partially transposed 
density matrix of the composite state remains positive in the course of the evolution. 
If the initial temperature of the thermal states is lower 
than a lower critical value $T_{lc}\leq T_{uc}$ the minimal 
eigenvalue of the partially transposed density matrix of the 
composite state becomes negative  which means that entanglement 
develops. We calculate the lower bound $T_{lb}$ for  $T_{lc}$ and show that  the negativity of 
the composite state is negligibly small in the interval $T_{lb}<T<T_{uc}$. 
Therefore the lower bound temperature $T_{lb}$ can be considered as \textit{the} critical 
temperature for the generation of entanglement.
\end{abstract}

\maketitle

\section{Introduction}
Efficient simulation of quantum dynamics on classical computers is hampered by the problem of scaling: 
the complexity of computation in quantum dynamics scales exponentially 
with the number of degrees of freedom \cite{Feynman}. The reason for this exponential growth 
is the entanglement of the degrees of freedom that is generated during the evolution. 
This  problem is of a fundamental character: entanglement is viewed as one of the main 
peculiarities of the quantum dynamics as compared to its classical counterpart \cite{Schroedinger,peres98}. 
Asking under what conditions entanglement is generated along the evolution of the quantum 
system is closely associated with the question of the quantum-classical transition \cite{Joos,Zurek}.

It is customary in quantum dynamical simulations to assume that the initial state 
of the composite system is factorized in the relevant local basis \cite{davis}. 
An important question is whether the product form is conserved along 
the evolution \cite{Lindblad,Gemmer1}. The answer was generally found 
to be negative both for the pure \cite{Gemmer1,Durt} and for the mixed state \cite{Durt} dynamics.

A pure composite state is entangled if and only if it is not factorized  in the local  basis. For mixed states the situation 
is more complex  \cite{Bruss}. For a bipartite composite system  
separability \cite{Werner} is defined as a decomposition of the density matrix of the composite 
system in the form 
\begin{eqnarray}
\Op \rho_{12}=\sum  p_i \Op \rho_1^i \otimes \Op \rho_2^i , \label{separability}
\end{eqnarray}
where $0\leq p_i \leq 1$ and $\sum_i p_i=1$  and $\Op \rho_1$ and $\Op \rho_2$ are density matrices on 
Hilbert spaces of the first and the second subsystem, respectively. Separable states  
exhibit only classical correlations. States that cannot be represented in the form (\ref{separability}) exhibit correlations that cannot
be explained within any classical theory  and are said to be entangled. There are two qualitatively different kinds of the mixed states entanglement \cite{horodecki}: free entanglement and bound entanglement. Free entanglement can be brought in a form useful for quantum information processing and bound entanglement is "useless" in this sense.

Separable states 
are not of the product form generally. Thus the important question remains, 
under what conditions the mixed state of the composite system evolving 
from the initial product (or generally separable ) state develops entanglement along the evolution. 
If quantum correlations in the composite system do not develop during the evolution  
one may speculate that the dynamics of the composite system  has   classical character. A possible practical implication is
that this "separable dynamics" could be simulated efficiently on classical computers.

\begin{figure}[t]
\epsfig{file=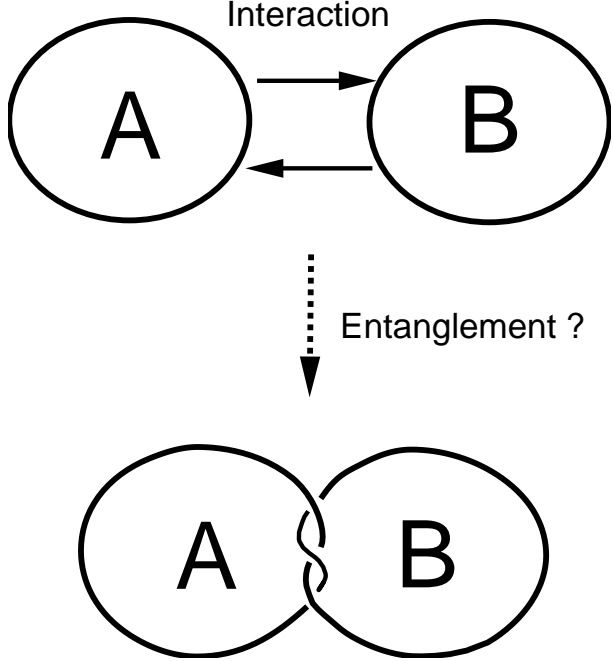, width=8.0cm, clip=} 
\caption{The coupling scheme for two directly interacting systems.}
\label{scheme:1}
\end{figure}

The dynamics of entanglement was investigated recently in various systems:  
the quantum Brownian particle \cite{Eisert}, harmonic chain \cite{Audenaert}, two-qubits system interacting with the common harmonic bath \cite{braun},
Jaynes-Cummings model \cite{Scheel}, NMR \cite{doronin}, various spin systems \cite{Hutton1,Hutton2,sen}, Morse oscillator coupled to the spin bath \cite{david}  and
bipartite Gaussian states in quantum optics \cite{Dodonov}  
to mention just some cases. 
The purpose of the present paper is to explore the temperature dependence of
entanglement generation  in the course of evolution of a bipartite state in the limit of weak coupling and nonresonant interaction between the parts.
Under these limitations nondegenerate perturbation theory was applied 
to the calculation of the bipartite entanglement in the evolving composite system. 
 We have considered two cases of interaction - (1) direct interaction, 
when two initially disentangled systems are brought into contact 
( Cf. Fig.\ref{scheme:1}), and 
(2) indirect interaction, when two noninteracting and initially disentangled systems  
are brought into contact with the third party   (Cf. Fig.\ref{scheme:2}). 
In each case the initial state of the composite system was taken 
to be the product of the thermal states of the parts.

To establish quantum entanglement the Peres-Horodeckii criterion is employed \cite{Peres,Horodeckii}.
The Peres-Horodeckii criterion  states that the bipartite system  is entangled when the partially transposed density matrix of the system
possesses a negative eigenvalue. The converse statement is generally not true: 
there exist inseparable states whose partially transposes are positive \cite{horodeckii}. 
It is proved in Ref.\cite{horodecki} that  states whose  partially transposes are positive (PPT states in what follows)
do not exhibit free entanglement. Therefore PPT states are either separable 
or bound entangled and as a consequence are not useful in quantum information processing.
In the context of simulating a quantum composite system with classical computers, we are interested in the 
possibility of maintaining a separable form  (Cf. Eq.(\ref{separability})) during the evolution.
We conjecture on the basis of Ref.\cite{Horodecki}, where it is proved that PPT density matrices of sufficiently small rank are separable, that for a state that remains PPT during the evolution separability can be
obtained by embedding in a larger Hilbert space.  

Applying the Peres-Horodeckii criterion to the case (1)  we show that for sufficiently low  initial 
temperature of the subsystems  the  interaction does induce entanglement unless the ground 
state of either one of the subsystem is invariant under the interaction. 
A lower and upper critical temperatures $T_{lc}$ and $T_{uc}$ exist such that 
if the composite system evolves from 
the initial thermal state  with temperature $T<T_{lc}$   the minimal eigenvalue of the 
partially transposed density matrix becomes negative in the course of the evolution  
and if $T>T_{uc}$   the minimal eigenvalue of the 
partially transposed density matrix  stays positive.  
The lower bound $T_{lb}$ of the lower critical temperatures $T_{lc}$  was calculated
in the  limit of weak intersystem coupling and shown to be tight: the negativity 
of the composite state \cite{vidal}, which is a quantitative counterpart of the Peres-Horodeckii criterion
and a measure of entanglement,  is shown to be generally negligible for temperatures 
in the interval  $T_{lb}<T<T_{uc}$. 
Therefore, according to the Peres-Horodeckii criterion, when $T<T_{lb}$ 
the composite system develops entanglement in the
course of the evolution and when $T>T_{lb}$ the composite state remains PPT state.  

The question addressed in case (2) of indirect coupling is  what are the conditions on the interaction 
with the common bath and on the initial temperature of the states which cause 
entanglement of the noninteracting systems? Two scenarios with time scales separation are studied: 
(a) two "slow" noninteracting systems coupled to a "fast" third party  (b) two "fast" noninteracting systems 
coupled to "slow" third party.  Under some technical assumptions about the form of the interaction 
we find in both cases that for sufficiently low initial temperature of the noninteracting 
systems entanglement  is induced by the interaction with the third party.  
We calculate  the lower bound temperature $T_{lb}$ in both cases of the time scales separation.
In the system of two noninteracting spins, coupled to the common bath,  the lower bound coincides with the $T_{uc}$.

In both cases (1) and (2) the evolution starts from an uncorrelated initial state of the 
composite system represented by the tensor product of thermal states of the subsystems involved.
As a consequence, initially the eigenstates of the partially transposed density matrix of the 
composite state are nonnegative. The evolution under the interaction perturbs the initial state. 
The new eigenvalues of the partially transposed density matrix are calculated by 
the nondegenerate perturbation theory assuming the coupling is weak and the interaction 
is nonresonanant.
The time dependence of the minimal eigenvalue is not analyzed in detail. 
As the time evolution of the density matrix is quasiperiodic the minimal 
eigenvalue of the partially transposed density matrix is also a  
quasiperiodic function of time. 
The interaction is said to induce entanglement if
the minimal eigenvalue becomes negative  in the course of the evolution.

\section{Entanglement between two directly interacting systems}

A composite system $A \otimes B $  evolves under the following Hamiltonian  :
\begin{eqnarray}
\Op H_{total} &=& \Op H+ \gamma \Op V.\label{hamiltonian1}
\end{eqnarray}%
where  $\Op H= \Op H_a\otimes \Op 1+\Op 1 \otimes \Op H_b$,  $\Op V= \Op V_a \otimes \Op V_b$ 
and $\gamma$ scales the magnitude of the interaction.
Let the initial state be 
\begin{eqnarray}
\Op \rho(0)&=& \Op \rho_a \otimes \Op \rho_b,\label{tensorproduct1}
\end{eqnarray}
where both $\Op \rho_a$ and $\Op \rho_b$ are thermal  states:
$\Op \rho_{a,b}=Z_{a,b}^{-1}\exp(- \Op H_{a,b}/T)$, where $Z_{a,b}^{-1}$ 
is the normalization factor. The Boltzman constant $k_B$  is  unity throughout the paper.
The evolution is followed in the interaction picture.  
 Then
\begin{eqnarray}
\frac{\partial \Op \rho'}{\partial t}&=&  -i \gamma \left[\Op V_a(t)\otimes \Op V_b(t), \Op \rho' \right],
\end{eqnarray}
where 
\begin{eqnarray}
\Op \rho'(t)&=&e^{-i \Op Ht} \Op \rho \ e^{i \Op H t},\nonumber \\
\Op V_{a,b}(t)&=&e^{i \Op H_{a,b}t} \Op V  e^{-i \Op H_{a,b}t}.
\end{eqnarray}
Here and in the rest of the paper we take $\hbar=1$. It is clear that  the density matrix $ \Op \rho (t)$ is separable if and only if $\Op \rho'(t)$ 
separable.
In what follows the tags in $\Op \rho'(t)$ are omitted for simplicity.
In the first order in the coupling $\gamma$ the evolution of $\Op \rho$ becomes:
\begin{eqnarray}
\Op \rho(t)&=& \Op \rho(0)- i\gamma \int_0^t \left[\Op V_a(t')\otimes 
\Op V_b(t'), \Op \rho(0) \right]dt'. \label{densityoperator}
\end{eqnarray}

Entanglment of $\Op \rho(t)$ is established by the application  
of the Peres-Horodeckii criterion. 
This is carried out by calculating the partial transpose of the state. The partial transposition $T_a$ 
with respect to subsystem $A$ of a bipartite state $\Op \rho_{ab}$ 
expanded in a local orthonormal basis as  
$\Op \rho_{ab}=\sum \rho_{ij,kl}\left|i \right\rangle \left\langle j \right|  \otimes \left|k \right\rangle \left\langle l \right| $ is defined as:
\begin{eqnarray}
\rho_{ab}^  {T_a}\equiv \sum \rho_{ij,kl}\left|j \right\rangle \left\langle i \right|  \otimes \left|k \right\rangle \left\langle l \right|.
\end{eqnarray}
The spectrum of the partially transposed density matrix does not depend on the choice of local basis or on the choice of the subsystem with respect to which the partial transposition is performed. By the Peres-Horodeckii criterion the eigenvalues of a partially transposed \textit{separable} bipartite state are nonnegative.

The density operator (\ref{densityoperator}) under the partial transposition ($T^a$) becomes:
\begin{eqnarray}
\Op \rho(t)^  {T_a}&=& \Op \rho(0)^  {T_a}- i\gamma \int_0^t \left[\Op V_a(t')\otimes \Op V_b(t'), 
\Op \rho(0) \right]^  {T_a}dt'.
\end{eqnarray}

Let $ \left|i k\right\rangle \equiv \left|i \right\rangle \otimes \left|k \right\rangle$ 
be the local orthonormal basis of the  
system $A \otimes B $ composed of the eigenvectors of the unperturbed  
Hamiltonian $\Op H= \Op H_a+ \Op H_b$:
\begin{eqnarray}
\Op H_{a,b}\left|i \right\rangle &=& E_{a,b}^i \left|i \right\rangle,
\end{eqnarray}
where $E_{a,b}^i$, $i=1,2,...$, is the unperturbed energy spectrum of the Hamiltonian $\Op H_{a,b}$. 
 The initial state is of the tensor product form, 
Cf. Eq.(\ref{tensorproduct1}), therefore :
 \begin{eqnarray}
\Op \rho(0)^  {T_a}\left|i k\right\rangle = \Op \rho(0)\left|i k \right\rangle = \Op \rho_a \otimes \Op \rho_b \left|i k\right\rangle=P_{i k}\left|i k \right\rangle , \label{eigenvectors0}
\end{eqnarray}
where $P_{i k}\equiv p_{a,i} p_{b,k}$ and $p_{a,i}, p_{b,k}$ are defined by 
$p_{a,i} =\langle i | \Op \rho_{a}|i \rangle$
and $p_{b,k} = \langle k | \Op \rho_{b}|k \rangle$.
The matrix elements of $\Op \rho(t)^  {T_a}$ in the chosen basis  are given by:
\begin{eqnarray}
\left\langle i k \left| \Op \rho(t)^  {T_a}  \right|j l\right\rangle = P_{i k} \delta_{(i k),(jl)} + M_{ik,jl} , \label{rhoelements}
\end{eqnarray}
where 

\begin{eqnarray}
M_{ik,jl} &=& i\gamma \int_0^t  \left\langle i k \left|(\left[\Op V_a(t')\otimes \Op V_b(t'), \Op \rho_a\otimes \Op \rho_b \right])^  {T_a}\right|j l\right\rangle dt' \nonumber \\
&=& i\gamma \int_0^t (\left\langle i k \left|  \Op \rho_a \Op V_a(t')^T \otimes \Op V_b(t')\Op \rho_b \right|j l\right\rangle \nonumber \\ 
&-& i\gamma \left\langle i k \left| \Op V_a(t')^T \Op \rho_a \otimes \Op \rho_b \Op V_b(t') \right|j l\right\rangle)dt' \label{secularmat} \\
&=& i\gamma (P_{il}-P_{jk})\int_0^t \left\langle i \left|\Op V_a(t')^T \right|j\right\rangle \left\langle k \left|\Op V_b(t') \right|l \right\rangle dt'\nonumber \\
&=& i\gamma (P_{il}-P_{jk})\int_0^t \left\langle j \left|\Op V_a \right|i \right\rangle \left\langle k \left|\Op V_b \right|l \right\rangle e^{i t'(E_a^i-E_a^{j}+E_b^l-E_b^{k})}dt'\nonumber \\
&=& \gamma (P_{il}-P_{jk}) \left\langle j \left|\Op V_a \right|i \right\rangle \left\langle k \left|\Op V_b \right|l\right\rangle \frac{e^{i t(E_a^i-E_a^{j}+E_b^l-E_b^{k})}-1}{(E_a^i-E_a^{j}+E_b^l-E_b^{k})}, \nonumber
\end{eqnarray}
where $\Op X^T$ designates the transpose of the operator $\Op X$.

When $T=0$, the zero eigenvalue of the initial state  $\Op \rho(0)$ is degenerate. 
As a result the zero eigenvalue of the partially transposed 
initial density operator $\Op \rho(0)^  {T_a}=\Op \rho(0)$ is also degenerate. 
The zero eigenvalues correspond to empty initially unoccupied states.
By the standard secular perturbation theory the first order correction to 
the degenerate eigenvalue $\lambda^{(0)}=0$ of the matrix $\Op \rho(0)^  {T_a}$ is given by
\begin{eqnarray}
\left|M_{n n'}-\lambda^{(1)}\delta_{n n'}\right|=0, \label{eigenproblem}
\end{eqnarray}
where $\left|n \right\rangle$ and $\left|n' \right\rangle$ are eigenvectors of the matrix $\Op \rho_a^T\otimes \Op \rho_b$,
corresponding to the degenerate $\lambda^{(0)}=0$.
Since $\Op \rho(0)^  {T_a}=\Op \rho_a^T\otimes \Op \rho_b=\Op \rho_a\otimes \Op \rho_b$  the eigenvectors of  $\Op \rho_a^T\otimes \Op \rho_b$ , corresponding to $\lambda^{(0)}=0$ are $\left|n \right\rangle =\left\{ \left|1 \right\rangle \otimes \left|i \right\rangle, \left|i \right\rangle \otimes \left|1 \right\rangle , \left|i \right\rangle \otimes \left|j\right\rangle,  i,j=2,3,..\right\}$. 
 
Therefore at $T=0$, $P_{i k}=\delta_{i k}\delta_{k1}$,  and by inspection of Eq. (\ref{secularmat}),
the only nonvanishing matrix elements in the degenerate subspace spanned by $\left|n \right\rangle$ 
and $\left|n' \right\rangle$ are $M_{1 i,j 1}$ and $M_{j 1, 1 i}$  where either $i\neq1$ or $j\neq1$.
Since the trace of the matrix $M$ is zero,  either all its eigenvalues vanish 
or some of them are negative. 
All the eigenvalues of $M$ cannot vanish unless $M=0$, which from Eq. (\ref{secularmat}) 
implies $\left[ \Op V_a,\Op \rho_a\right]=0$ or $\left[ \Op V_b,\Op \rho_b\right]=0$, i.e. 
the  ground state of either one of the subsystem is invariant under the interaction. 
In this case the interaction acts locally on the subsystems and  cannot entangle them. 
Otherwise there are negative  solutions to Eq. (\ref{eigenproblem}) and 
as a consequence the partial transpose of the density operator attains negative eigenvalues 
already in the first order in the coupling. Therefore, according to the Peres-Horodeckii 
criterion, entanglement  develops  at zero temperature.

To simplify the study of the generation of entanglement at finite temperatures it is assumed that 
the only non zero matrix elements of $\Op V_{a,b}$ are those between neighboring states, 
i.e. $(\Op V_{a,b})_{ij}\propto \delta _{i,j\pm 1}$. 
Under this assumption the partially transposed density matrix $\Op \rho(t)^  {T_a}$ obtains 
the following structure:
\begin{eqnarray}
 \Op \rho(t)^  {T_a}=
\begin{pmatrix} 
 P_{11} & 0 & 0 & M_{11,22} & 0 & 0 & 0 & . & .  \\
 0 & P_{12} & M_{12,21} & 0 & M_{12,23} & 0 & 0 & . & .  \\
 0 & M_{12,21}^* &P_{21} & 0 & 0 & M_{21,32} & 0 & . & .  \\
 M_{11,22}^* & 0 & 0 & P_{22} & 0 & 0 & M_{22,33} & . & .  \\
 0 & M_{12,23}^* & 0 & 0 & P_{23} & M_{23,32} & 0 & . & .  \\
 0 & 0 & M_{21,32}^* & 0 & M_{23,32}^* & P_{32} & 0 & . & .  \\
 0 & 0 & 0 & M_{22,33}^* & 0 & 0 & P_{33} & . & .  \\
 . & . & . & . & . & . & . & . & .  \\
 .& . & . & . & . & .& . & . & .  \\
\end{pmatrix} , \label{matrixform}
\end{eqnarray}
where $P_{ij}$ are defined after the Eq.(\ref{eigenvectors0}) and 
$M_{ki,jl}$ by Eq.(\ref{secularmat}). 

There are two kinds of matrix elements $M_{ki,jl}$:
$M_{ki,(k+1) (i+1)}$ and $M_{ki,(k\pm1) (i\mp1)}$ (other elements are their counterparts under the transposition). Matrix elements $M_{ki,(k+1) (i+1)}$ 
couple the unperturbed eigenvalues $P_{ki}$ and $P_{(k+1) (i+1)}$.
For small coupling strength $\gamma$ $\left|M_{ki,(k+1) (i+1)}\right|\ll P_{ki}$ 
and the contribution of $M_{ki,(k+1) (i+1)}$ to the correction to $P_{ki}$ is negligible 
and cannot make the  eigenvalue negative. 
On the other hand, the ratio 
$\left|M_{ki,(k+1) (i+1)}\right|/ P_{(k+1) (i+1)}\propto \gamma (P_{k (i+1)}-P_{(k+1)i})/P_{(k+1) (i+1)}$ 
can in general be arbitrary large for low temperatures but for sufficiently high temperatures 
it tends to zero and as a consequence the contribution of the coupling element 
$M_{ki,(k+1) (i+1)}$ to the correction to $P_{(k+1)(i+1)}$ is negligible.
It can be checked along the same lines that the ratio of the coupling matrix elements  
$M_{ki,(k\pm1) (i\mp1)}$ to the  unperturbed eigenvalues $P_{ki}$ and
$P_{(k\pm1) (i\mp1)}$ of the partially transposed density matrix 
(\ref{matrixform}) vanish for sufficiently high temperature.  
Therefore, at least for composite systems with finite Hilbert space dimensions, 
there exists a finite upper critical temperature $T_{uc}$. Above  $T_{uc}$ the spectrum 
of the partially transposed density matrix remains positive (PPT).  In close vicinity of $T_{uc}$ from below 
the minimal eigenvalue becomes negative in the course of the evolution. These conclusions stay in accord with a general result \cite{gurvitz,bandyopadhyay} that  finite dimensional composite states  in sufficiently small neighbourhood of the maximally mixed state (i.e. thermal states at infinite temperature) are separable.
We conjecture, that for an  infinite composite system,  
the upper critical temperature $T_{uc}$ exists if the energy spacing is bound. 

At sufficiently low initial temperature the minimal eigenvalue of the partially transposed 
density matrix becomes negative in the course of the evolution. This means that there exists a finite 
lower critical temperature $T_{lc}$. Below $T_{lc}$ the composite systems $A\otimes B$ 
develops entanglement. In sufficiently close vicinity of $T_{lc}$ from above the 
state  remains PPT in the course of evolution. It is possible that $T_{lc} = T_{uc}$. This equality
is confirmed  in all numerical tests. 
A lower bound $T_{lb}$ for the lower critical temperature $T_{lc}$ can be calculated using perturbation analysis. 
It is shown that this bound is tight since the free entanglement in the interval $T_{lb}<T<T_{uc}$ 
is negligibly small under the weak coupling assumption. Therefore, from the practical point 
of view the lower bound $T_{lb}$ for $T_{lc}$ can be considered as \textit{the} 
critical temperature for entanglement. For simplicity the lower bound  
for the lower critical temperature is termed "the lower bound  temperature" throughout the paper.

At low temperatures the leading order contribution to the negative eigenvalue of the partially transposed 
density matrix comes from the matrix elements $M_{11,22}$, $M_{12,21}$ 
(and their complex conjugates) that do not vanish at $T=0$.
Therefore, to the leading order in $\gamma$, the nonvanishing eigenvalues of the  
partially transposed density matrix Eq.(\ref{matrixform}) are the eigenvalues of the following 
effective partially transposed density 
matrix $\Op \rho(t)^  {T_a}_{eff}$ :
\begin{eqnarray}
 \Op \rho(t)^  {T_a}_{eff}=
\begin{pmatrix} 
 P_{11} & 0 & 0 & M_{11,22}   \\
 0 & P_{12} & M_{12,21} & 0   \\
 0 & M_{12,21}^* &P_{21} & 0   \\
 M_{11,22}^* & 0 & 0 & P_{22} 
\end{pmatrix} . \label{matrixform4}
\end{eqnarray}
The critical temperature, 
calculated for the effective $4\times4$ matrix (\ref{matrixform4}), is  
a lower bound for the lower critical temperature $T_{lc}$ of the bipartite system $A\otimes B$. 
The eigenvalues of Eq. (\ref{matrixform4}) are eigenvalues of two $2\times2$ matrices:
\begin{eqnarray}
\left(
\begin{array}{clrr} 
 P_{12} & M_{12,21}  \\
 M_{12,21}^* &P_{21} 
\end{array} \right) \label{twotwo}
\end{eqnarray}
and
\begin{eqnarray}
\left(
\begin{array}{clrr} 
 P_{11} & M_{11,22}  \\
 M_{11,22}^* &P_{22} 
\end{array} \right). \label{twotwo1}
\end{eqnarray}
The eigenvalues of the matrix (\ref{twotwo}) are:
\begin{eqnarray}
\lambda_{\pm} =\frac{ P_{12}+P_{21}\pm \sqrt{(P_{12}+P_{21})^2-4(P_{12}P_{21}-|M_{12,21}|^2)}}{2},\label{thelambda} 
\end{eqnarray}
where from Eq.(\ref{secularmat}):
\begin{eqnarray}
 M_{12,21}=\gamma  \left\langle 2 \left|\Op V_a \right|1 \right\rangle \left\langle 2 \left|\Op V_b \right|1 \right\rangle \frac{e^{i t\Delta E_{11}}-1}{\Delta E_{11}}
 (P_{11}-P_{22}), 
 \label{eq:a12}
\end{eqnarray}
where we define  $\Delta E_{11} = E_a^2-E_a^{1}+E_b^2-E_b^{1}$, which is the lowest 
joint excitation energy of  the composite system.

From Eq. (\ref{thelambda}), $\lambda_{-}$ will be negative whenever 
$P_{12}P_{21}<|M_{12,21}|^2$ and positive if $P_{12}P_{21}>|M_{12,21}|^2$. 
The lower bound temperature $T_{lb}$ is evaluated from the condition $P_{12}P_{21}=|M_{12,21}|^2$. 
Since $|M_{12,21}|$ is an oscillating function of time (Cf. Eq. (\ref{eq:a12}) )
the amplitude of $|M_{12,21}|$ is taken to be equal to $\sqrt{P_{12}P_{21}}$:
\begin{eqnarray}
\frac{2 \gamma }{\Delta E_{11}}\left| \left\langle 2 \left|\Op V_a \right|1 \right\rangle \left\langle 2 \left|\Op V_b \right|1 \right\rangle \right|(P_{11}-P_{22})
=\sqrt{P_{12}P_{21}}. 
\end{eqnarray} 
Assuming  that $T_{lb}$ is low  $P_{11}-P_{22}\approx P_{11}$ and then
\begin{eqnarray}
\frac{2 \gamma }{\Delta E_{11}}\left| \left\langle 2 \left|\Op V_a \right|1 \right\rangle \left\langle 2 \left|\Op V_b \right|1 \right\rangle \right|
=\sqrt{\frac{P_2P_3}{P_1^2}}=e^{-\frac{\Delta E_{11}}{2 T_{lb}}}. 
\end{eqnarray} 
Since $e^{-\frac{\Delta E_{11}}{2 T}}$ is a monotonic function of the temperature, 
at $T>T_{lb}$ $\lambda_{-}>0$ and at $T<T_{lb}$ $\lambda_{-}<0$.
Finally, the expression for the lower bound temperature $T_{lb}$ becomes:
\begin{eqnarray}
T_{lb}=-\frac{\Delta E_{11}}{2\ln\left(\frac{2 \gamma }{\Delta E_{11}}\left| \left\langle 2 \left|\Op V_a \right|1 \right\rangle \left\langle 2 \left|\Op V_b \right|1 \right\rangle \right|\right)
}. \label{temperature}
\end{eqnarray}
 
 So far only two of the eigenvalues of the matrix (\ref{matrixform4}) have been evaluated. 
The other two  eigenvalues are found to be strictly positive at and above the  temperature  $T_{lb}$.
Therefore, the expression (\ref{temperature}) defines the critical temperature for the partially 
transposed effective density matrix (\ref{matrixform4}) and the lower bound temperature  of the partially 
transposed density matrix (\ref{matrixform}).

\begin{figure}[t]
\epsfig{file=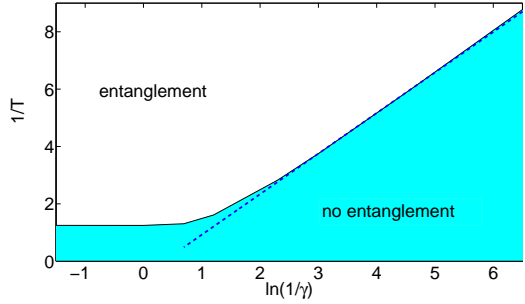, width=8.0cm, clip=} 
\caption{The shaded area in the parameter space of the inverse 
initial temperature $T$ of two spins and the logarithm of the inverse coupling strength 
$\gamma$, represents values of $T$ and $\gamma$, where entanglement  does not develop 
in the course of the evolution. The composite system of two spins evolves
from the initial product of thermal states under 
the Hamiltonian 
$\Op H=\frac{1}{2}\omega(\Op \sigma_z^a\otimes \Op 1+ (\sqrt{2}-1)\Op 1\otimes \Op \sigma_z^b)+\gamma 
( \Op \sigma_x^a \otimes \Op \sigma_x^b- \Op \sigma_y^a \otimes \Op \sigma_y^b)$. 
The evolution is calculated numerically for $\omega=1$. 
The border of the shaded area represents $T_{uc}$ calculated numerically.
The dashed line represents $T_{lb}$ according to Eq. (\ref{temperature}). 
Up to the coupling $\gamma=0.1$ $T_{lb}$ approximates $T_{lc}$ very well.}
\label{fig:1}
\end{figure}

Eq. (\ref{temperature}) can be generalized to an interaction term of the form 
$\sum \gamma_i \Op V_a^i\otimes \Op V_b^i$:

\begin{eqnarray}
T_{lb}=-\frac{\Delta E_{11}}{2\ln\left(\frac{2  }{\Delta E_{11}}\left|\sum_i \gamma_i \left\langle 2 \left|\Op V_a^i \right|1 \right\rangle \left\langle 2 \left|\Op V_b^i \right|1 \right\rangle \right|\right)
}, \label{temperaturegen}
\end{eqnarray}
provided $\sum_i \gamma_i \left\langle 2 \left|\Op V_a^i \right|1 \right\rangle \left\langle 2 \left|\Op V_b^i \right|1 \right\rangle \neq 0$. When this term vanishes there is no entanglement in the first order in the coupling strength $\gamma$.

For the system of two interacting spins the lower bound $T_{lb}$ given by Eq. (\ref{temperature})
coincides with the upper critical temperature $T_{uc}$ therefore in this case \textit{the} critical temperature exists in the strict sense.
Fig. \ref{fig:1} shows results of the numerical calculation of the critical temperature as a function of coupling strength for a system 
of two interacting spins evolving from the initial product state of two thermal states. 
The Peres-Horodeckii criterion was used and the partial transpose of the evolving density matrix 
was calculated numerically to determine entanglement. The shaded area in the parametric space of  
the logarithm of inverse coupling and the inverse initial temperature represents the values of the parameters 
where no entanglement develops. 
For coupling up to $\gamma=0.1$ $T_{lb}$ given by Eq. (\ref{temperature}) (the dashed line) 
corresponds well to the numerical values of $T_{uc}$. It is interesting to note that 
for large values of coupling the critical temperature  asymptotically tends to a 
finite constant value of the same order of magnitude as the energy difference between 
the first excited and the ground state of the unperturbed composite system.

At $T<T_{lb}$ the minimal eigenvalue of the partially transposed state 
(\ref{matrixform})
 is negative.  
We want to show that above $T_{lb}$ the negative eigenvalues of the matrix (\ref{matrixform})  are  of higher order in $\gamma$ and therefore are negligibly small
when the coupling is weak.

 Let's consider corrections to the eigenvalues $P_{i(j+1)}$ and $P_{(i+1)j}$ of the composite state  (\ref{matrixform}). 
The order of magnitude estimate of the smallest  one of the corrected eigenvalues is : $\lambda_{-}^{ij}=\sqrt{P_{i(j+1)}P_{(i+1)j}}-\gamma P_{ij}/\Delta E_{ij}$, where $\Delta E_{ij} \equiv E_a^{i+1}-E_a^{i}+E_b^{j+1}-E_b^{j}$. For simplicity we assume $P_{i(j+1)}=P_{(i+1)j}$. Then $\lambda_{-}^{ij}=O(P_{i(j+1)}-\gamma P_{ij}/\Delta E_{ij})$. Below $T_{lb}$ the minimal eigenvalue of the state  (\ref{matrixform}) is $\lambda_{-}=
O(-\gamma /\Delta E_{11})$. We shall estimate the ratio $r^{ij}\equiv \lambda_{-}^{ij}/\lambda_{-}$ and show that it is negligible when the coupling is weak. 
We shall assume without loss of generality that the ground state energy is zero: $E_a^{1}+E_b^{1}=0$. Then the partition function $Z$ of the composite system is larger than unity. It follows that

\begin{eqnarray}
r^{ij}=\frac{\lambda_{-}^{ij}}{\lambda_{-}}=\frac{\frac{\gamma}{\Delta E_{ij}} P_{ij}-P_{i(j+1)}}{\frac{\gamma}{\Delta E_{11}} }< \frac{\frac{\gamma}{\Delta E_{ij}} Z P_{ij}-Z P_{i(j+1)}}{\frac{\gamma}{\Delta E_{11}} }=\frac{\frac{\gamma}{\Delta E_{ij}} e^{-E_{ij}/T}-e^{- E_{i(j+1)}/T}}{\frac{\gamma}{\Delta E_{11}} }.\label{inequality}
\end{eqnarray}
 We are looking for the maximal value of $r^{ij}$ in the interval  $0<T<T_c^{ij}$, corresponding to the condition  $P_{i(j+1)}<\gamma P_{ij}/\Delta E_{ij}$, i.e. to the negative values of $\lambda_{-}^{ij}$. $T_c^{ij}$ is determined by the condition $\lambda_{-}^{ij}=0$. The ratio $r^{ij}$ is positive in the interval $0<T<T_c^{ij}$ and vanishes  on its borders. Therefore  $r^{ij}$ has a maximum $r_m^{ij}$ at $0<T_m^{ij}<T_c^{ij}$, which is found from the condition  $\partial r^{ij}/\partial T|_{T_m^{ij}}=0$. The calculation gives $\exp(-\Delta E_{ij}/(2T_m^{ij}))=(\gamma/\Delta E_{ij})(E_{ij}/E_{i(j+1)})<(\gamma/\Delta E_{ij})= \exp(-\Delta E_{ij}/(2T_c^{ij}))$, which proves that there is one maximum $r_m^{ij}$  at $0<T_m^{ij}<T_c^{ij}$. We remark, that $T_c^{ij}$, corresponding to the largest $\Delta E_{ij}$ over all $i$ and $j$, $T_{uc}^*$, is of the order of the upper critial temperature $T_{uc}^*=O(T_{uc})$. The maximal value of $r^{ij}$ is given by:

\begin{eqnarray}
r_m^{ij}=\frac{\Delta E_{11}}{2E_{ij}+\Delta E_{ij}}\left(\frac{2E_{ij}}{2E_{ij}+\Delta E_{ij}}\right)^{\frac{2E_{ij}}{\Delta E_{ij}}}\left(\frac{\gamma}{\Delta E_{ij}}\right)^{\frac{2E_{ij}}{\Delta E_{ij}}}<\frac{\Delta E_{11}}{2E_{ij}+\Delta E_{ij}}\left(\frac{\gamma}{\Delta E_{ij}}\right)^{\frac{2E_{ij}}{\Delta E_{ij}}},
\end{eqnarray}
where the inequality follows from the fact that $1/e<\left(\frac{2E_{ij}}{2E_{ij}+\Delta E_{ij}}\right)^{\frac{2E_{ij}}{\Delta E_{ij}}}<1$ in general. As a next step we notice that $\Delta E_{11}\leq 2E_{ij}$, therefore
\begin{eqnarray}
r^{ij}<r_m^{ij}<\frac{\Delta E_{11}}{2E_{ij}+\Delta E_{ij}}\left(\frac{\gamma}{\Delta E_{ij}}\right)^{\frac{2E_{ij}}{\Delta E_{ij}}}\leq \frac{\Delta E_{11}}{\Delta E_{11}+\Delta E_{ij}}\left(\frac{\gamma}{\Delta E_{ij}}\right)^{\frac{\Delta E_{11}}{\Delta E_{ij}}}.
\end{eqnarray}
Introducing the definition $x_{ij}\equiv \Delta E_{ij}/\Delta E_{11}$ and taking $\Delta E_{11}=1$, which corresponds to a rescaling of the coupling strength $\gamma$, leads to:
\begin{eqnarray}
\frac{\lambda_{-}^{ij}}{\lambda_{-}}\equiv r^{ij}<\frac{1}{x_{ij}}\left(\frac{\gamma}{x_{ij}}\right)^{\frac{1}{x_{ij}}}. \label{scaling}
\end{eqnarray}

Typically the spectrum becomes denser with increasing energy.
In that case $x_{ij}\equiv\Delta E_{ij}/\Delta E_{11}\leq 1$. Values of  $\lambda_{-}^{ij}$, corresponding to $x_{ij}\ll1$ need not be taken into account, because $T_c^{ij}<T_{lb}$ in this case and as a conseqence  $\lambda_{-}^{ij}>0$ at $T\geq T_{lb}$. At $x_{ij}=O(1)$ the upper bound for $r$ scales as $O(\gamma)$ and therefore corresponding negative eigenvalues of Eq. (\ref{matrixform}) are negligible. In this case we expect that $T_{lb}\approx T_{uc}$.

In those cases when $x_{ij}\equiv\Delta E_{ij}/\Delta E_{11}\gg 1$ the upper bound for $r$ scales as $O(1/x_{ij})$ and the corresponding negative eigenvalues of Eq. (\ref{matrixform}) can be neglected, too.

When $x_{ij}$ is moderately larger than unity the upper bound Eq.(\ref{scaling}) for $r^{ij}$ has a local maximum. The position of the maximum weakly depends on $\gamma$: numerical calculations show $x_{ij}\approx 2-10$ in the range of $10^{-4}\leq\gamma\leq10^{-1}$ . The value of the minimum is a monotonically slowly increasing function of $\gamma$. In the range $10^{-4}\leq\gamma\leq10^{-1}$ numerical estimation of Eq.(\ref{scaling}) shows  values $0.04-0.1$ for the local maximum. It is clear that the upper bound Eq.(\ref{scaling}) for $r^{ij}$ is far from being tight. In fact, numerical calculations show that $r^{ij}$ is generally much smaller. As a consequence, the corresponding negative eigenvalues of Eq. (\ref{matrixform}) can be neglected.

It can be argued that although each one of the negative eigenvalues of Eq. (\ref{matrixform}) is negligible at $T\geq T_{lb}$ the (free) entanglement of the state cannot
be neglected. 
In fact, the minimal negative eigenvalue of the partially transposed matrix is not a  measure of entanglement. 
Various measures of entanglement have been defined \cite{virmani}. In the present context we will
employ a quantitative counterpart of the Peres-Horodeckii criterion, the negativity \cite{vidal}:
\begin{eqnarray}
N(\Op \rho(t))\equiv \frac{\left\|\Op \rho(t)^  {T_a}\right\|-1}{2},
\end{eqnarray}
where $\left\|\Op X \right\|= Tr\sqrt{\Op X^{\dagger}\Op X}$ is the trace norm of an operator $\Op X$.
The negativity of the  state equals the absolute value of the sum of  the negative eigenvalues 
of the partially transposed state. When the negativity of a composite bipartite state  vanishes  
there is no free entanglement in the state.
It can be shown by the order of magnitude analysis similar to the analysis above that values of the negativity of the composite state, corresponding to the partial transpose  (\ref{matrixform}), are generally dominated by the minimal negative eigenvalue. As a consequence, the negativity of the state, evolving from the initial thermal product state at the temperature $T\geq T_{lb}$, is negligible under the weak coupling assumption.

\begin{figure}[t]
\epsfig{file=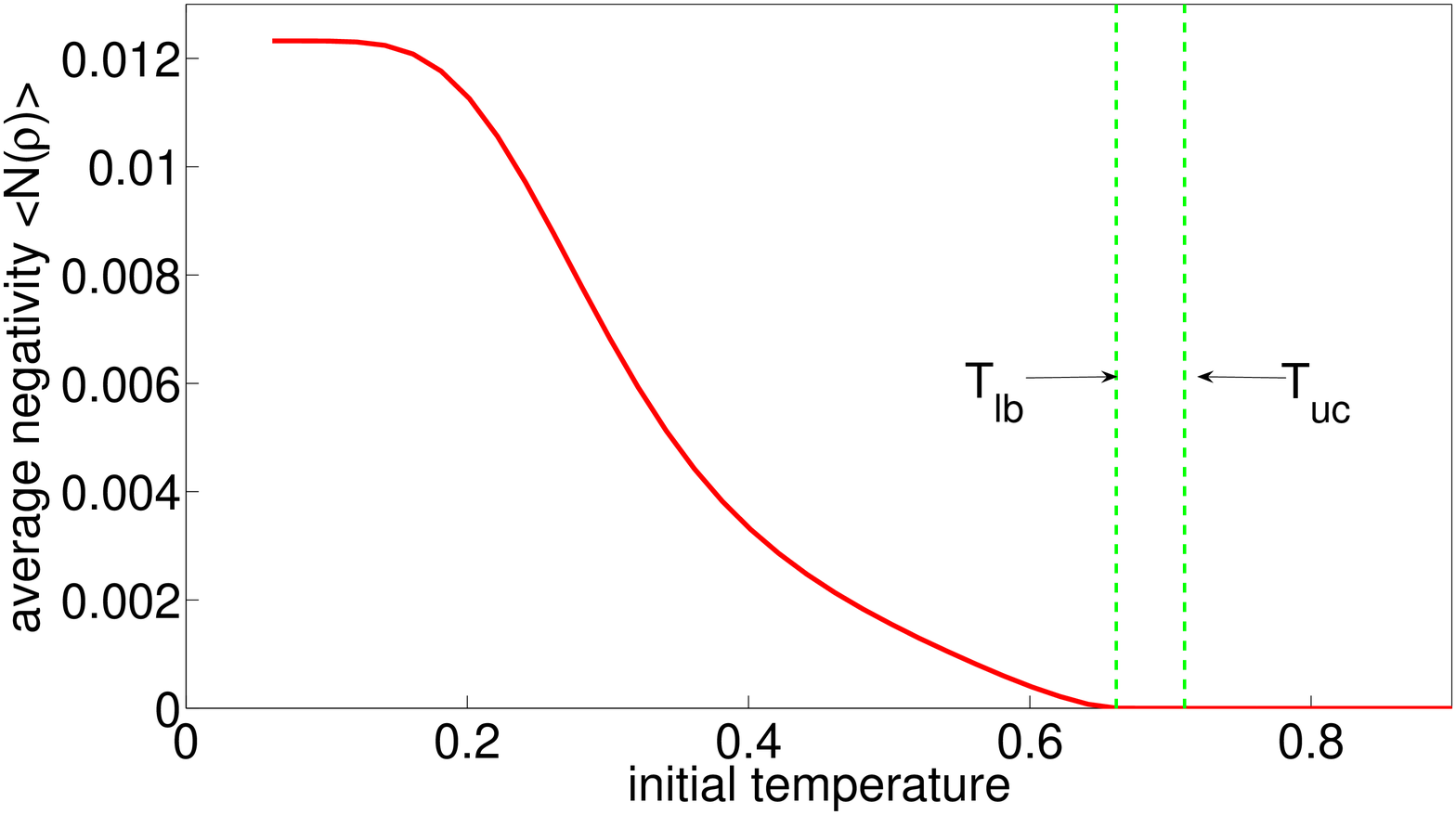, width=8.0cm, clip=} 
\caption{The time averaged negativity as a function of initial temperature.  
The composite system is constructed from two interacting four level subsystems. The initial state
is a product of thermal states. The evolution is generated numerically by the Hamiltonian 
(\ref{hamiltonian1}) (for details of the Hamiltonian see the text) with $\gamma=0.05$. 
The dashed lines correspond to the lower bound temperature $T_{lb}$,  
Eq. (\ref{temperature}), and to the upper critical temperature $T_{uc}$,  found numerically. 
It can be seen that the entanglement is vanishingly small in the interval $T_{lb}<T<T_{uc}$.}
\label{fig:7}
\end{figure}

\begin{figure}[t]
\epsfig{file=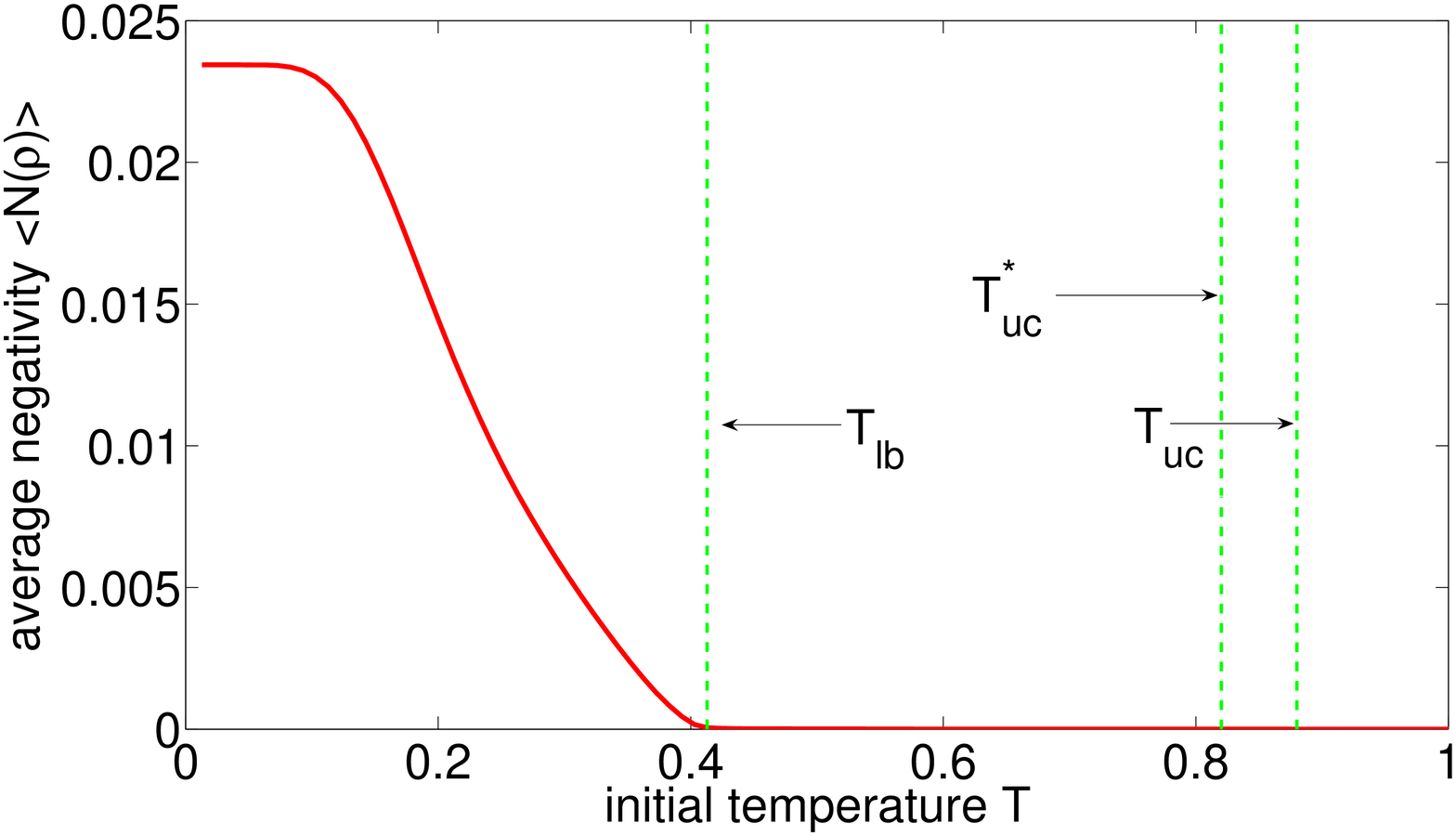, width=8.0cm, clip=} 
\caption{The time averaged negativity as a function of initial temperature of the 
composite system.  The composite system is constructed from two interacting four level subsystems. The initial state
is a product of thermal states. The evolution is generated numerically by the Hamiltonian 
(\ref{hamiltonian1}) (for details of the Hamiltonian see the text) with $\gamma=0.05$.
The dashed lines correspond to the lower bound temperature $T_{lb}$ Eq. (\ref{temperature}),  
to the numerical value of the upper critical temperature $T_{uc}$ and to the value $T_{uc}^*$, corresponding  
to the largest spectrum spacing $\Delta E_{max}$. 
We see that entanglement is vanishingly small at $T_{lb}<T<T_{uc}$, as expected, and $T_{uc}^*$ 
is a good approximation to the upper critical temperature $T_{uc}$.}
\label{fig:6}
\end{figure}

Figures \ref{fig:7} and \ref{fig:6} display results of numerical calculations of the time averaged negativity of the composite state (\ref{densityoperator}) as a function of initial temperature for two different kinds of unperturbed spectra of the composite system $A\otimes B$.  
Both $A$ and $B$ are four level systems. The composite system evolves from the initial  product 
of thermal states of $A$ and $B$  under the Hamiltonian (\ref{hamiltonian1}).

 Fig. \ref{fig:7} presents the results of calculations for the following choice of the unperturbed spectra of $\Op H_a$ and $\Op H_b$:  $E_a^{\left\{1,2,3,4\right\}}=\left\{1,5,8,10\right\}$ and   
$E_b^i=\sqrt{E_a^i}$. Care was taken to avoid resonances and the spectra were chosen to become denser with increasing energy. The interaction terms in the 
Hamiltonian were restricted to $(\Op V_{a,b})_{ij}=\delta _{i,j\pm 1}$ and the coupling 
strength $\gamma=0.05$. We see that $T_{uc} \approx T_{lb}$ and the time averaged negativity 
$\left\langle N(\Op \rho(t))\right\rangle$ is negligible in the interval $T_{lb}<T<T_{uc}$ as expected.

Fig. \ref{fig:6} displays the time averaged negativity 
$\left\langle N(\Op \rho(t))\right\rangle$ as a function of initial temperature of the  
composite state of two interacting four level subsystems $A$ and $B$ 
with the unperturbed energy spectra   $E_a^{\left\{1,2,3,4\right\}}=\left\{1,3,7,13\right\}$ and   $E_b^i=\sqrt{E_a^i}$. 
The composite state evolves from the initial  product of two thermal states under 
the Hamiltonian (\ref{hamiltonian1}), where $(\Op V_{a,b})_{ij}=\delta _{i,j\pm 1}$ 
and the coupling strength $\gamma=0.05$. In choosing the unperturbed spectra  
care was taken to avoid resonances and to ensure that the maximal value of
$x_{ij}\equiv \Delta E_{ij}/\Delta E_{11}$ equals the position of the local maximum of the upper bound (\ref{scaling}), corresponding to $\gamma=0.05$.  Fig. \ref{fig:7} shows that 
the time averaged negativity $\left\langle N(\Op \rho(t))\right\rangle$ is negligible in the interval 
$T_{lb}<T<T_{uc}$ as expected. The value of $T_c^{ij}\equiv T_{uc}^*$ (the definition of $T_c^{ij}$ is given after Eq.(\ref{inequality})), corresponding to the maximal value $\Delta E_{max}\equiv \max_{ij}(\Delta E_{ij})$ is calculated. $T_{uc}^*$ 
is in good correspondence with the value $T_{uc}$, calculated numerically.

\section{Entanglement between two noninteracting systems in contact with a common third party}
\begin{figure}[t]
\epsfig{file=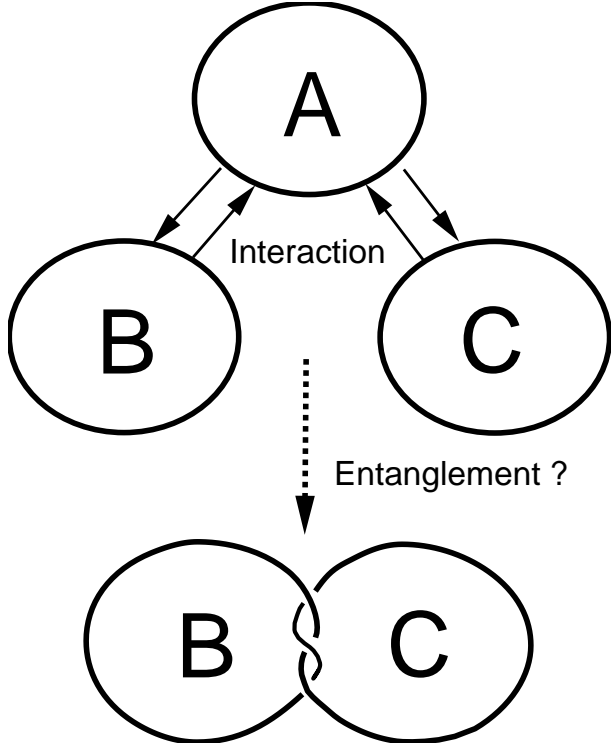, width=8.0cm, clip=} 
\caption{Scheme of interaction for two noninteracting systems in contact with a common third party.}
\label{scheme:2}
\end{figure}
The dynamics studied is of the composite system $A\otimes B\otimes C$ where systems $B$ and $C$   
do not interact directly (Cf. Fig.\ref{scheme:2}). The entanglement explored is of the reduced composite 
system  $B\otimes C$.

The evolution is generated by the following Hamiltonian:
\begin{eqnarray}
\Op H_{total}=\Op H+\gamma\Op V,
\end{eqnarray}
where
$ \Op H= \Op H_a+ \Op H_b+ \Op H_c$ and
$\Op V=\gamma \Op V_a\otimes ( \Op V_b\otimes \Op 1+ \Op 1\otimes \Op V_c)$. The analysis is carried out in the interaction picture. 
The initial state is taken to be $\Op \rho(0)=\Op \rho_a\otimes\Op \rho_b\otimes\Op \rho_c$, where $\Op \rho_a$, $\Op \rho_b$ and $\Op \rho_c$ are thermal states. Since
$B$ and $C$ are noninteracting entanglement will appear only in the second order in the coupling. 
Up to second order in $\gamma$ the state of the composite system
$A\otimes B\otimes C$ becomes:

\begin{eqnarray}
\Op \rho(t)'&=& \Op \rho(0)- i\gamma \int_0^t \left[\Op V(t'), \Op \rho(0)' \right]dt'\nonumber \\
&-&\gamma^2 \int_0^t \int_0^{t'}\left[\Op V(t'),\left[\Op V(t''), \Op \rho(0)'\right] \right]dt'dt'',
\end{eqnarray}
where 
\begin{eqnarray}
\Op \rho'=e^{-i \Op H t}\Op \rho \ e^{i \Op H t},\nonumber \\
\Op V(t)=e^{i \Op H t} \Op V  e^{-i \Op H t}.
\end{eqnarray}
In what follows the tag above the  $\Op \rho(t)$ is omitted.

Next the system is reduced to $B\otimes C$ by taking the partial trace of $\Op \rho(t)$ over the system $A$ 
degrees of freedom and the partial transposition with respect to the subsystem $B$ is taken:
\begin{eqnarray}
\Op \rho_{bc}^  {T_b}(t)&=& \Op \rho_{bc}^  {T_b}(0)+ \Op M(t), \label{rhobc}
\end{eqnarray}
where $\Op \rho_{bc}(t)\equiv Tr_a (\Op \rho(t))$ and
\begin{eqnarray}
\Op M&\equiv& - i\gamma \int_0^t Tr_a\left(\left[\Op V(t'), \Op \rho(0) \right]\right)^  {T_b}dt'\nonumber \\
&-&\gamma^2 \int_0^t \int_0^{t'}Tr_a \left( \left[\Op V(t'),\left[\Op V(t''), \Op \rho(0)\right] \right] \right)^  {T_b}dt'dt''.\label{secondorderdensity}
\end{eqnarray}
 
Let $ \left|i k\right\rangle \equiv \left|i \right\rangle \otimes \left|k \right\rangle$ 
be the local orthonormal basis of the  
system $B\otimes C $ composed of the eigenstates of the  Hamiltonian $\Op H_b+\Op H_c$:
\begin{eqnarray}
\Op H_{b,c}\left|i \right\rangle &=& E_{b,c}^i \left|i \right\rangle,
\end{eqnarray}
where $E_{b,c}^i$, $i=1,2,...$, is the unperturbed energy spectrum of the Hamiltonian $\Op H_{b,c}$. 
Since $\Op \rho_{bc}(0)= \Op \rho_b\otimes\Op \rho_c$:
 \begin{eqnarray}
\Op \rho_{bc}(0)^  {T_b}\left|i k\right\rangle = \Op \rho_{bc}(0)\left|i k \right\rangle = \Op \rho_b \otimes \Op \rho_c \left|i k\right\rangle=P_{i k}\left|i k \right\rangle , 
\end{eqnarray}
 where $P_{i k}\equiv p_{b,i} p_{c,k}$, and $p_{a,i}, p_{b,k}$ are defined by 
$p_{b,i} =\langle i | \Op \rho_{a}|i \rangle$
and $p_{c,k} = \langle k | \Op \rho_{b}|k \rangle$.
 The matrix elements of $\Op \rho_{bc}(t)^  {T_b}$ are given by:
\begin{eqnarray}
\left\langle i k \left| \Op \rho_{bc}(t)^  {T_b}  \right|j l\right\rangle = P_{i k} \delta_{(i k),(jl)} + M_{ik,jl} , \label{rhoelements1}
\end{eqnarray}
where by definition $M_{ik,jl}=\left\langle i k \left|\Op M\right|j l\right\rangle$. 

From this point the calculations proceed along the same lines as in Section II following 
Eq.(\ref{rhoelements}). The minimal eigenvalue of the partially transposed reduced 
state $\Op \rho_{bc}(t)^  {T_b}$ is shown to be negative at sufficiently low temperatures 
and the lower bound temperature $T_{lb}$
is calculated.

The negative eigenvalue of the partially transposed composite state Eq.(\ref{rhobc})
is calculated  to the leading order in the coupling strength $\gamma$ assuming  
$\left\langle n_{i}\left|\Op V_{i}\right|m_{i}\right\rangle\propto \delta _{n_i \ m_i\pm1}$. 
As in Section II the eigenvalue is found from the spectrum of the  $2\times 2$ matrix:
\begin{eqnarray}
\left(
\begin{array}{clrr} 
 P_{12}+M_{12,12} & M_{12,21}  \\
 M_{12,21}^* &P_{21}+M_{21,21}
\end{array} \right), \label{twotwo2}
\end{eqnarray}
completely analogous to the matrix  (\ref{twotwo}). 
The eigenvalues of Eq. (\ref{twotwo2}) are:
\begin{eqnarray}
\lambda_{\pm} &=&\frac{ P_{12}+M_{12,12}+P_{21}+M_{21,21}}{2} \label{thenextlambda} \\
&\pm& \frac{\sqrt{(P_{12}+M_{12,12}+P_{21}+M_{21,21})^2-4((P_{12}+M_{12,12})(P_{21}+M_{21,21})-|M_{12,21}|^2)}}{2}\nonumber  
\end{eqnarray}
and the  eigenvalue $\lambda_{-}$  becomes negative when $(P_{12}+M_{12,12})(P_{21}+M_{21,21})<|M_{12,21}|^2$.

To calculate $M_{12,12}$, $M_{21,21}$ and $M_{12,21}$ we first note that
  the integrand in the first order term in  Eq. (\ref{secondorderdensity}) is:
\begin{eqnarray}
&Tr_a& \left(\left[\Op V(t'), \Op \rho(0) \right]\right)^{T_b}= \left\langle \Op V_a\right\rangle \left[ \Op V_{bc}(t'),\Op \rho_{bc}(0) \right]^{T_b} \nonumber \\
&=&\left\langle \Op V_a\right\rangle \left(\left[\Op V_b(t'), \Op \rho_b\right]^T\otimes \Op \rho_c 
+ \Op \rho_b^T\otimes \left[ \Op V_c(t'),\Op \rho_c \right] \right)  \label{firstorder} \\
&=&-\left\langle \Op V_a\right\rangle\left(\left[\Op V_b(t')^T, \Op \rho_b\right]\otimes \Op \rho_c 
-  \Op \rho_b\otimes \left[ \Op V_c(t'),\Op \rho_c \right]\right),\nonumber 
\end{eqnarray}
where $\left\langle \Op V_a \right\rangle$ means the thermal average of the operator $\Op V_a$ and the notation $\Op V_{bc}\equiv \Op V_b\otimes \Op 1+ \Op 1\otimes \Op V_c$ is introduced. The initial condition $\Op \rho_{bc}(0)=\Op \rho_b\otimes\Op \rho_c$ was used. Since $\Op \rho_{b,c}\left|i \right\rangle = \delta_{i,1}\left|i \right\rangle$ the term Eq. (\ref{firstorder}) does not contribute to the eigenvalues of the matrix (\ref{twotwo2}) in the first order.

To simplify the calculation of the second order corrections  it is assumed  that the thermal average of the system $A$ coupling operator $\left\langle \Op V_a \right\rangle$ vanishes. This assumption is not crucial for the qualitative picture of temperature dependence of the entanglement.
Moreover, it is in line with common models of coupling, for example,  
the Caldeira-Leggett model \cite{Caldeira}, 
dipole interaction with the electromagnetic field \cite{Carmichael}, etc. 
The integrand in the second order term in Eq. (\ref{secondorderdensity}) is:
\begin{eqnarray}
&Tr_a& \left(\left[\Op V(t'),\left[\Op V(t''), \Op \rho(0)\right] \right]\right)^{T_b}=\left\langle \Op V_a(t') \Op V_a(t'')\right\rangle \left[ \Op V_{bc}(t'),\Op V_{bc}(t'') \Op \rho_{bc} \right]^{T_b}\nonumber \\
&-&\left\langle \Op V_a(t'') \Op V_a(t')\right\rangle \left[ \Op V_{bc}(t'), \Op \rho_{bc} \Op V_{bc}(t'')\right]^{T_b}.
\end{eqnarray}
Expanding the thermal averages in the orthonormal basis $\left| n \right\rangle$ of the Hamiltonian $H_a$ 
leads to:
\begin{widetext}
\begin{picture}(3.375,0)
\end{picture}
\begin{eqnarray}
Tr_a \left( \left[\Op V(t'),\left[\Op V(t''), \Op \rho(0)\right] \right]\right)^{T_b}
&=& \sum_{m,n}p_{a,n}\left|\left\langle m\left| \Op V_a\right|n\right\rangle\right|^2 (\cos(\omega_{m n}^a(t'-t''))\left[ \Op V_{bc}(t'),\left[\Op V_{bc}(t''), \Op \rho_{bc} \right]\right]\nonumber \\
&+&i\sin(\omega_{m n}^a(t'-t''))\left[ \Op V_{bc}(t'),\left\{\Op V_{bc}(t''), \Op \rho_{bc} \right\}\right])^{T_b}
\label{secondorder},
\end{eqnarray}
\hfill
\begin{picture}(3.375,0)
  \put(0,0){\line(1,0){3.375}}
  \put(0,0){\line(0,-1){0.08}}
\end{picture}
\end{widetext} 
where $\omega_{m n}^a$ is the energy difference between the states $\left| n \right\rangle$ and  $\left| m \right\rangle$ of the Hamiltonian $H_a$, the $\left\{\Op X,\Op Y\right\}$ designates anticommutator of operators $\Op X$ and $\Op Y$ and $p_{a,n}\equiv (\Op \rho_a)_{n n}$.

For simplicity  the notation $\Op C(t',t'')$ is used for the operator (\ref{secondorder}).
Expressing the operator $\Op V_{bc}$ in terms of $\Op V_b$ and $\Op V_c$ we put the matrix elements of  $\Op C(t',t'')$ into the following form:
\begin{widetext}
\begin{picture}(3.375,0)
\end{picture}
\begin{eqnarray}
\left\langle 1 2\right|\Op C(t',t'')\left| 1 2 \right\rangle
&=& -2P_{11}\sum_{m,n}p_{a,n}\left|\left\langle m\left| \Op V_a\right|n\right\rangle\right|^2 \left|\left\langle 1\left|\Op V_c\right|2\right\rangle\right|^2 \cos((\omega_{m n}^a+\omega_{c})(t'-t''))\nonumber \\
\left\langle  2 1\right|\Op C(t',t'')\left|  2 1\right\rangle
&=& -2P_{11}\sum_{m,n}p_{a,n}\left|\left\langle m\left| \Op V_a\right|n\right\rangle\right|^2 \left|\left\langle 1\left|\Op V_b\right|2\right\rangle\right|^2 \cos((\omega_{m n}^a+\omega_{b})(t'-t''))\label{elements} \\
\left\langle 1 2\right|\Op C(t',t'')\left|  2 1 \right\rangle&=&P_{11} \sum_{m,n}p_{a,n}\left|\left\langle m\left| \Op V_a\right|n\right\rangle\right|^2 \left\langle 2\left|\Op V_b\right|1\right\rangle \left\langle 2\left|\Op V_c\right|1\right\rangle  e^{i\omega_{m n}^a(t'-t'')}(e^{-i(\omega_{b}t'+\omega_{c}t'')} \nonumber \\ &+& e^{-i(\omega_{b}t'' +\omega_{c}t')})
,\nonumber
\end{eqnarray}
\hfill
\begin{picture}(3.375,0)
  \put(0,0){\line(1,0){3.375}}
  \put(0,0){\line(0,-1){0.08}}
\end{picture}
\end{widetext} 
where
$\omega_{b,c}$ stands for the energy difference between the first excited  and the ground states of the unperturbed subsystem $B$ ( $C$).
The matrix elements  $M_{12,12}$, $M_{21,21}$ and $M_{12,21}$ are given by:
\begin{eqnarray}
M_{12,12}&=&-\gamma^2 \int_0^t \int_0^ {t'}\left\langle 1 2\right|C(t',t'')\left| 1 2 \right\rangle dt'dt''.\nonumber \\
M_{21,21}&=&-\gamma^2 \int_0^t \int_0^{t'}\left\langle 21\right|C(t',t'')\left| 21 \right\rangle dt'dt'' \label{matrix}\\
M_{12,21}&=&-\gamma^2 \int_0^t \int_0^{t'}\left\langle 1 2\right|C(t',t'')\left| 21 \right\rangle dt'dt''.\nonumber 
\end{eqnarray}

The integration is straightforward but the final expressions are cumbersome. Two cases are considered explicitly: $(a)$ $\omega_{m n}^a>>\omega_{b,c}$ and $(b)$ $\omega_{b,c}>>\omega_{m n}^a$. 
In both cases it is shown that at sufficiently low initial temperature of the system $B \otimes C$ one of the eigenvalues of the  matrix  (\ref{twotwo2}) is negative and the lower bound temperature $T_{lb}$ is calculated. 
 
\subsection{Two "slow" systems interacting with a "fast" common third party}
Performing the integrations in Eq. (\ref{matrix}) and taking the leading terms in $\omega_{b,c}/\omega_{m n}^a$ brings to:
\begin{widetext}
\begin{picture}(3.375,0)
\end{picture}
\begin{eqnarray}
M_{12,12}
&=& 4\gamma^2\sum_{m,n}p_{a,n}\left|\left\langle m\left| \Op V_a\right|n\right\rangle\right|^2 \left|\left\langle 1\left|\Op V_c\right|2\right\rangle\right|^2 
\frac{\sin((\omega_{m n}^a+\omega_c)t/2)^2}{(\omega_{m n}^a)^2}\nonumber \\
M_{21,21}
&=& 4\gamma^2\sum_{m,n}p_{a,n}\left|\left\langle m\left| \Op V_a\right|n\right\rangle\right|^2 \left|\left\langle 1\left|\Op V_b\right|2\right\rangle\right|^2 
\frac{\sin((\omega_{m n}^a+\omega_b)t/2)^2}{(\omega_{m n}^a)^2} \label{matrixexplicite}\\
M_{12,21}
&=& 2 \gamma^2\sum_{m,n}p_{a,n}\left|\left\langle m\left| \Op V_a\right|n\right\rangle\right|^2 \left\langle 2\left|\Op V_b\right|1\right\rangle \left\langle 2\left|\Op V_c\right|1\right\rangle \frac{(1- e^{-i(\omega_b+\omega_c)t})}{\omega_{m n}^a(\omega_b+\omega_c)}
.\nonumber
\end{eqnarray}
\hfill
\begin{picture}(3.375,0)
  \put(0,0){\line(1,0){3.375}}
  \put(0,0){\line(0,-1){0.08}}
\end{picture}
\end{widetext} 

At $T=0$ the minimal eigenvalue of Eq. (\ref{thenextlambda}) is given by $\lambda_{-}=-\sqrt{M_{12,12}M_{21,21}-|M_{12,21}|^2}$, which
 to the leading order in $\omega_{b,c}/\omega_{m n}^a$ gives $\lambda_{-}=-|M_{12,21}|^2$ . 
This proves that the system $B\otimes C$ becomes entangled at sufficiently low temperature. 
We note that this result holds at any finite temperature of the system $A$. At infinite temperature of the system $A$ $M_{12,21}\equiv 0$ and no free entanglement is generated in the system $B\otimes C$.

At finite initial temperature of $B\otimes C$ the condition  $\lambda_{-}<0$ translates to 
$P_{12}P_{21}<\gamma^4 |M_{12,21}|^2P_{11}^2$ 
to the leading order in $\omega_{1,2}/\omega_{m n}^a$. 
The lower bound temperature $T_{lb}$ is found from the condition $P_{12}P_{21}=\gamma^4 |M_{12,21}|^2P_{11}^2$. Since 
$|M_{12,21}|$ is an oscillating function of time  the amplitude of $|M_{12,21}|$ must be substituted for $|M_{12,21}|$ 
in this equality, which leads to the following equation defining the lower bound temperature:
\begin{eqnarray}
&4\gamma^2&\frac{\left| \left\langle 2\left|\Op V_b\right|1\right\rangle \left\langle 2\left|\Op V_c\right|1\right\rangle \right|}{\omega_b+\omega_c}\sum_{m,n}\frac{p_{a,n}\left|\left\langle m\left| \Op V_a\right|n\right\rangle\right|^2 }{\omega_{m n}^a} = \sqrt{\frac{P_2P_3}{P_1^2}}\nonumber \\
&=&\exp({-\frac{\omega_b+\omega_c}{2 T_{lb}}}),
\end{eqnarray}
finally leading to:
\begin{eqnarray}
T_{lb} =  \frac{-(\omega_b+\omega_c)}{2\ln\left(4\gamma^2\frac{\left| \left\langle 2\left|\Op V_b\right|1\right\rangle \left\langle 2\left|\Op V_c\right|1\right\rangle \right|}{\omega_b+\omega_c}\sum_{m,n}\frac{p_{a,n}\left|\left\langle m\left| \Op V_a\right|n\right\rangle\right|^2 }{\omega_{m n}^a} \right)}.\label{temperature2}
\end{eqnarray}
A generalization of the formula to the case of  interaction of the form 
$\sum \gamma_i \Op V_a^i\otimes( \Op V_{b}^i\otimes \Op 1+ \Op 1\otimes \Op V_{c}^i)$ 
can be carried out along the same lines.

\begin{figure}[t]
\epsfig{file=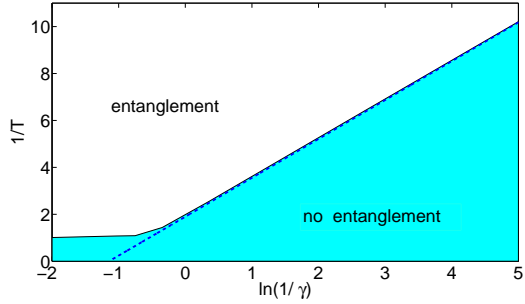, width=8.0cm, clip=} \caption{
The shaded area in the parameter space of the inverse 
initial temperature $T$ of the "slow" spins and the logarithm of the inverse coupling strength 
$\gamma$, represents values of $T$ and $\gamma$ where entanglement  does not develop 
in the course of the evolution. The composite system of two "slow" spins interacting 
with a "fast" four level system evolves from the initial product of thermal states under 
the Hamiltonian (\ref{figure2}). The dashed line is the plot of $T_{lb}$,  Eq.  (\ref{temperature2}). 
Up to the coupling $\gamma=1$ its correspondence to the border of the shaded area is very good.}
\label{fig:2}
\end{figure}

The entanglement in the reduced system of two noninteracting "slow" 
spins interacting with the "fast" four level "bath"
was explored numerically and the results are plotted on Fig. \ref{fig:2}. 
The shaded area in the parametric space  of the logarithm of inverse coupling strength
and the inverse initial temperature of the spins represents parametric 
values for which no entanglement develops in the course of the evolution. 
The border of the shaded area corresponds to the critical temperature for various coupling magnitudes.
The Hamiltonian of the composite system is:
\begin{eqnarray}
\Op H&=& \Op H_a \otimes \Op 1_b\otimes \Op 1_c 
 + \frac{1}{2}\omega (\Op 1_a\otimes(\Op \sigma_z^b\otimes \Op 1_c+\sqrt2 \Op 1_b\otimes \Op \sigma_z^c))  \label{figure2} \\
&+& \gamma \Op V_a \otimes (\Op \sigma_x^b\otimes \Op 1_c+\Op 1_b\otimes \Op \sigma_x^c), \nonumber
\end{eqnarray}
where $(\Op H_a)_{ij}=\delta_{ij}E_a^i$, $E_a^{\left\{1,2,3,4\right\}}= \left\{0,10 \omega,20 \omega,30 \omega\right\}$ and $(\Op V_a)_{ij}=\delta_{ij}$.
The temperature of the thermal initial state of the "bath"  is  $T=5 \omega$. The value of $\omega$ chosen for the numerical calculation is unity.
The correspondence of  Eq. (\ref{temperature2}) (the dashed line) 
to the numerical values is very good up to a coupling strength of the order of unity. 
We note that for large values of the coupling strength $\gamma$ the critical temperature asymptotically tends to 
a finite constant value.

\subsection{Two "fast" systems interacting with a "slow" common third party}
The case $\omega_{b,c}>>\omega_{m n}^a$ is  more complex. 
To demonstrate entanglement at zero temperature of the system $B\otimes C$ two 
simplifying assumptions were added. The first is  that the temperature of the system 
$A$ is also zero.  The second is that the matrix elements of $V_a$ couple only the neighboring states: 
$\left\langle n\right|\Op V_a\left|m\right\rangle\propto \delta_{n,m\pm 1}$. 
Under these two assumptions the expressions for $M_{12,12}$, $M_{21,21}$ and  
$M_{12,12}M_{M_{21,21}}-|M_{12,21}|^2$ become:
\begin{eqnarray}
M_{12,12}&=&P_{11}\left(\frac{2\gamma \left|\left\langle 2\left| \Op V_a\right|1\right\rangle\right| \left|\left\langle 1\left|\Op V_c\right|2\right\rangle\right|\sin\left(\frac{(\omega_a+\omega_c)t}{2}\right)}{\omega_c}\right)^2 \nonumber \\
M_{21,21}&=&P_{11}\left(\frac{2\gamma \left|\left\langle 2\left| \Op V_a\right|1\right\rangle\right| \left|\left\langle 1\left|\Op V_b\right|2\right\rangle\right|\sin\left(\frac{(\omega_a+\omega_b)t}{2}\right)}{\omega_b}\right)^2 \nonumber \\ \nonumber \\
M_{12,12}M_{M_{21,21}}-|M_{12,21}|^2 \label{expression}  \\
&=& P_{11}^2\left(\frac{2\gamma^2\left|\left\langle 2\left| \Op V_a\right|1\right\rangle\right|^2 \left|\left\langle 1\left|\Op V_c\right|2\right\rangle\right| \left|\left\langle 1\left|\Op V_b\right|2\right\rangle\right|}{\omega_b\omega_c}\right)^2 S(t), \nonumber
\end{eqnarray}
where
\begin{eqnarray}
S(t)&=&\sin(\omega_at)[\sin(\omega_b t)+\sin(\omega_c t)\nonumber \\
&-&\sin((\omega_a+\omega_b + \omega_c)t)].
\end{eqnarray}

To estimate $S(t)$  new variables $x=\sin(\omega_at)$, $y=\sin(\omega_b t)$ and $z=\sin(\omega_c t)$ 
are introduced. 
Ignoring the zero measure set of commensurable frequencies  we can treat the function $S(t)$ as 
function of three independent variables $x$, $y$ and $z$.  The range of $S(t)$ in the cube, 
defined by $-1\leq x,y,z\leq1$, can be explored numerically and is found to be:
$s\leq S(t)\leq3$, where $s\approx-1.6834$. Therefore, from Eq.(\ref{expression})  
$M_{12,12}M_{M_{21,21}}-|M_{12,21}|^2<0$, which  proves that at zero temperature 
$\lambda_{-}<0$ (Cf. Eq.(\ref{thenextlambda})) and the systems $B$ and $C$ are 
entangled by the interaction with the system $A$.

The lower bound  temperature is determined by the condition $\lambda_{-}=0$, 
which translates to  $(P_{12}+M_{12,12})(P_{21}+M_{21,21})=|M_{12,21}|^2$ 
(Cf. Eq.(\ref{thenextlambda})). The latter condition can be put in the 
form $(M_{12,12}M_{21,21}-|M_{12,21}|^2)+P_{12}P_{21}+P_{12}M_{21,21}+P_{21}M_{12,12}=0$ . 
Since $M_{12,21}$ and $M_{21,21}$ are nonnegative independent functions of time  
the minimum value of $(M_{12,12}M_{21,21}-|M_{12,21}|^2)+P_{12}P_{21}+P_{12}M_{21,21}+P_{21}M_{12,12}$ 
is obtained at $M_{21,21}=M_{12,12}=0$. Then the lower  bound temperature can be calculated 
from the condition that the amplitude of $M_{12,12}M_{21,21}-|M_{12,21}|^2$ equals $-P_{12}P_{21}$:
\begin{eqnarray}
\frac{2\gamma^2\sqrt{\left|s\right|}\left|\left\langle 2\left| \Op V_a\right|1\right\rangle\right|^2 \left|\left\langle 1\left|\Op V_c\right|2\right\rangle\right| \left|\left\langle 1\left|\Op V_b\right|2\right\rangle\right|}{\omega_b\omega_c} = \sqrt{\frac{P_{12}P_{21}}{P_{11}^2}},
\end{eqnarray}
finally leading to:
\begin{eqnarray}
T_{lb} =  \frac{-(\omega_b+\omega_c)}{2\ln\left(2\gamma^2 \sqrt{\left|s\right|} \frac{\left|\left\langle 2\left| \Op V_a\right|1\right\rangle\right|^2 \left|\left\langle 1\left|\Op V_c\right|2\right\rangle\right| \left|\left\langle 1\left|\Op V_b\right|2\right\rangle\right|}{\omega_b\omega_c} \right)}.\label{temperature3}
\end{eqnarray}
It is interesting to note that $T_{lb}$ in this case does not depend on the time scales of the "slow" system.

\begin{figure}[t]
\epsfig{file=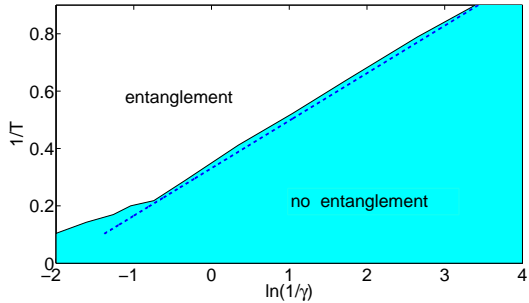, width=8.0cm, clip=} 
\caption{The shaded area in the parameter space of the inverse 
initial temperature $T$ of the "fast" spins and the logarithm of the inverse coupling strength 
$\gamma$, represents values of $T$ and $\gamma$ where entanglement  does not develop 
in the course of the evolution. The composite system of two "fast" spins interacting 
with the "slow" four level system evolves from the initial product of thermal states under 
the Hamiltonian (\ref{figure3}). The dashed line is the plot of $T_{lb}$,  Eq.  (\ref{temperature3}). 
Up to the coupling $\gamma=1$ its correspondence to the border of the shaded area is good.}
\label{fig:3}
\end{figure}

The entanglement in the reduced system of two noninteracting "fast" 
spins interacting with the "slow" four level  "bath" 
was explored numerically and the results are plotted on Fig. \ref{fig:3}. 
The shaded area in the parametric space  of the logarithm of inverse coupling strength
and the inverse initial temperature of the  spins represents parametric 
values for which no entanglement develops in the course of the evolution. 
The border of the shaded area corresponds to the critical temperature for various coupling magnitudes.
The Hamiltonian is chosen to be similar to the previous example, Cf. Eq.(\ref{figure2}), but time scales of the subsystems are reversed:
\begin{eqnarray}
\Op H&=& \Op H_a \otimes \Op 1_b\otimes \Op 1_c 
 + 5\omega (\Op 1_a\otimes(\Op \sigma_z^b\otimes \Op 1_c+\sqrt2 \Op 1_b\otimes \Op \sigma_z^c))  \label{figure3} \\
&+& \gamma \Op V_a \otimes (\Op \sigma_x^b\otimes \Op 1_c+\Op 1_b\otimes \Op \sigma_x^c), \nonumber
\end{eqnarray}
where $(\Op H_a)_{ij}=\delta_{ij}E_a^i$, $E_a^{\left\{1,2,3,4\right\}}=  \left\{0,\omega,2\omega,3\omega \right\}$ and $(\Op V_a)_{ij}=\delta_{ij}$.
The temperature of the thermal initial state of the "bath"  was chosen as $T=0.01\omega$, 
which is small compared to the energy scale of the "bath" chosen for the numerical calculation:
$\omega=1$.  The dashed line on the Fig. \ref{fig:3} is a plot of Eq. (\ref{temperature3})  and the correspondence to the border of the shaded area at coupling strength up to the order of unity is  good. 

\section{Summary an Conclusions}

Entanglement is created by both direct and indirect weak interaction between two initially disentangled systems 
prepared in thermal states at sufficiently low temperatures.  
The study is restricted to the conditions where the ground states of both systems are not invariant under the interaction 
and  the interaction is nonresonant. As a consequence, the present analysis left out some interesting models such 
as the Jaynes-Cummings model \cite{Jaynes}. The Jaynes-Cummings model   of interacting two level system and a quantized field mode 
was investigated in Ref.\cite{Scheel}. It was found that no free entanglement is generated 
in the course of the evolution of the composite system if the initial temperature of both the subsystems 
is sufficiently high. 

The generation of entanglement in cases of the weak \textit{resonant} direct and undirect interactions will be treated separately \cite{khasin}.

In the case of indirect interaction to show entanglement at $T=0$ we have assumed that the thermal average 
of the third party coupling term 
in the initial state vanishes. The reason for the assumption was  technical. It should be noted that many  
system-bath models of linear coupling satisfy this assumption \cite{Caldeira}. 
The additional technical assumption was that the coupling terms of the 
noninteracting parties possess matrix elements only between the adjacent energy states. 
Here, too, the assumption is  general  for weak coupling models. Two cases of 
time scale separation were considered explicitly. 
The first is the case of two "slow" systems interacting via the "fast" third common party.  
The second is the case of two "fast" systems interacting via the "slow" third common party.  
In the first case the entanglement was shown to appear at sufficiently low initial temperature of the "slow" 
systems for any finite temperature of the third party. In the second case  the entanglement  
develops at sufficiently low  initial temperature of the "fast" systems. In this case we assumed 
that the third party was prepared at zero temperature and that the third party coupling agent has nonvanishing 
matrix elements  only between the adjacent energy states. This assumption is stronger than just assuming 
that its thermal average vanishes. 

In  these  cases of indirect interaction and in the case of the direct interaction between the parts we have shown that if the initial temperature of the bipartite state is zero
entanglement is generated by the interaction. 
At sufficiently high temperature the composite state remains PPT in the course of evolution. 
From these results it follows that a lower critical temperature $T_{lc}$ exists: if the initial temperature  
of both thermal states is below $T_{lc}$   the interaction generates entanglement in the course 
of the evolution, and if the initial temperature is sufficiently close to $T_{lc}$ from above the  
the  composite state remains PPT forever. When the composite system is finite dimensional 
there exists an upper critical temperature $T_{uc}$: if the initial temperature  of both 
thermal states is higher than $T_{uc}$  the composite state remains PPT  in the course of evolution
and if the initial temperature is sufficiently close to $T_{uc}$ from below  entanglement is generated. 
We conjecture on the basis of numerical experiments that $T_{lc}=T_{uc}$ in general. 
In both cases of 
a direct and an indirect interaction between the initially disentangled systems, prepared in 
thermal states, we calculated the lower bound $T_{lb}$ for the lower critical temperature $T_{lc}$.  
When the initial temperature of both thermal states is below $T_{lb}$  the interaction generates entanglement 
in the course of the evolution. For temperatures above the lower bound $T_{lb}$ the negativity of the partially 
transposed composite state is zero in the leading order in the coupling strength and therefore negligible 
in the weak coupling limit. It follows, that $T_{lb}$ may be considered as the \textit{physical} 
critical temperature for the negativity of the 
composite state. 

Separable states can be considered as classical states, because they lack quantum correlations. 
One may hope  that, as a consequence, the dynamics of separable states can be efficiently simulated 
on classical computers. Whether this is possible is an open question in quantum information science. 
If a moderate scaling  procedure exists for the  simulation of the dynamics of a separable bipartite state,
then it seems that such a procedure exists also if the evolving state remains PPT for all times.  
Ref.\cite{Horodecki} has proved that a  density operator $\Op \rho$ supported on a $M\times N$ 
dimensional Hilbert space ($M\leq N$) with positive partial transpose  and a rank smaller than or equal 
to $N$ is separable. It follows that a PPT state of dimension $N$ is always separable when 
embedded in the larger Hilbert space of  dimension $N^2$ or higher. The dynamics of the low dimensional 
PPT state will be physically equivalent to the dynamics 
of the  high dimensional separable state which can (hopefully)  
be simulated efficiently on the classical computer. 

The present analysis shows that above a critical temperature $T_{lb}$ the PPT character of a composite state 
is preserved along the evolution. The challenge is to construct an effective simulation  
for the dynamics of  a composite quantum systems at finite temperature employing classically based computers.

\begin{acknowledgments} We want to dedicate this study to Assher Peres who passed away on February 2005. We are grateful to Roi Baer and Jose Palao for critical comments. This work is supported by DIP and the Israel Science Foundation.
\end{acknowledgments}

\end{document}